\documentclass[twocolumn,prl,superscriptaddress,longbibliography, english]{revtex4-1}
\usepackage{float,graphicx,amsmath,amssymb,bm}
\usepackage[colorlinks=true, citecolor=blue, urlcolor=blue, linkcolor=red]{hyperref}
\renewcommand{\section}[1]{{\par\it #1.---}\ignorespaces}

\begin{document}
\title{Tunable strong magnetic anisotropy in two-dimensional van der Waals antiferromagnets}
\author{Feiping Xiao}
\affiliation{School of Physics and Electronics, Hunan University, Changsha 410082, China}
\author{Qingjun Tong}
\email{tongqj@hnu.edu.cn} \affiliation{School of Physics and Electronics, Hunan University, Changsha 410082, China}
\date{\today}

\begin{abstract}
We show that anisotropic energy of a 2D antiferromagnet is greatly enhanced via stacking on a magnetic substrate layer, arising from the sublattice-dependent interlayer magnetic interaction that defines an effective anisotropic energy. Interestingly, this effective energy couples strongly with the interlayer stacking order and the magnetic order of the substrate layer, providing unique mechanical and magnetic means to control the antiferromagnetic order. These two types of control methods affect distinctly the sublattice magnetization dynamics, with a change of the ratio of sublattice precession amplitudes in the former and its chirality in the later. In moir\'{e} superlattices formed by a relative twist or strain between the layers, the coupling with stacking order introduces a landscape of effective anisotropic energy across the moir\'{e}, which can be utilized to create nonuniform antiferromagnetic textures featuring periodically localized low-energy magnons.

KEYWORDS: 2D magnets, antiferromagnetic order, van der Waals heterostructure, moir\'{e} pattern
\end{abstract}

\maketitle

\section{\color{blue}{Introduction}}\label{sec:introduction}
The demand of down scale of information carriers has spurred tremendous interest in two-dimensional (2D) layered magnets, because of their potential applications in constructing atomically thin spintronic devices \cite{Burch2018,Gong2019,Mak2019,Huang2020,Rahman2021}. Compared with 2D ferromagnets \cite{Gong2017,Huang2017,Jiang2018,Huang2018,Wang2018,Deng2018,Song2018,Klein2018,Wang2018m,Cenker2021}, antiferromagnetic (AFM) materials are more robust for information storage, owing to their immunity against external magnetic disturbance \cite{Wang2016,Du2016,Lee2016,Kim2019,Kang20202,Wang20212,Ni20212,Hwangbo2021}. In addition, AFM magnets possess two degenerate elementary chargeless bosonic modes (magnons), operating at terahertz regime and allowing for ultrafast transfer of spin information \cite{Kampfrath2011,Cheng2014,Rezende2019,Xing2019,Li2020}. A prerequisite for 2D AFM application is the existence of a strong and tunable magnetic anisotropy \cite{Jungwirth2016,Jungwirth2018}. First, a strong magnetic anisotropy is needed to stabilize 2D magnetism against thermal fluctuations \cite{Mermin1966}, and achieve terahertz AFM resonance frequency \cite{Keffer1952}. Second, a tunable magnetic anisotropy can be utilized to manipulate AFM order, whose control is usually quite difficult due to the absence of net magnetization. This motivates the ongoing searching for 2D AFM magnets with tunable strong magnetic anisotropy.

\begin{figure}
\centering\includegraphics[width=0.45\textwidth]{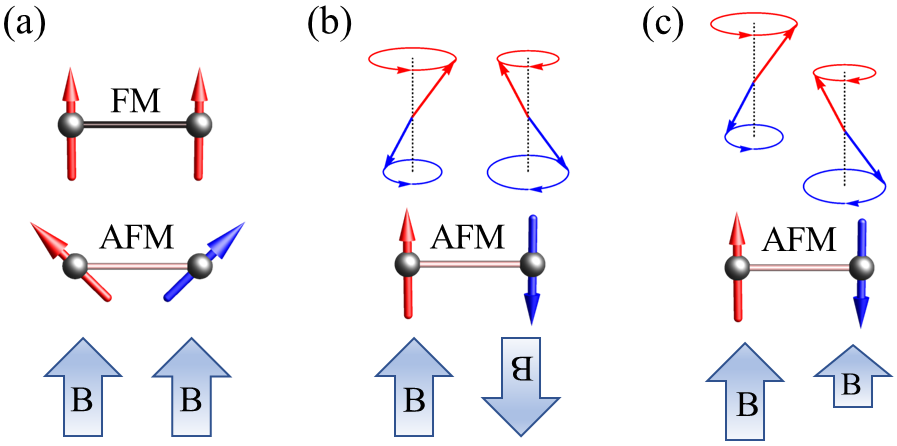}\caption{Schematics of the response of magnetic orders with weak uniaxial anisotropy to different sublattice magnetic field. (a) A uniform sublattice magnetic field stabilizes the FM order, while destabilizes the AFM order into spin flop phase. (b) A staggered sublattice magnetic field pointing in opposite direction stabilizes the AFM order with two degenerate AFM resonance modes. (c) A general staggered sublattice magnetic field stabilizes the AFM order with two nondegenerate AFM resonance modes.
\label{fig1}}
\end{figure}

Besides the material-determined intrinsic anisotropy, magnetic order in a ferromagnet can also be stabilized and oriented by an external magnetic field, achieving the spin polarized state (c.f. Figure \ref{fig1}a). In contrast, because of its staggered nature, the AFM magnetic order can only be controlled by a staggered sublattice magnetic field that points along opposite direction (c.f. Figure \ref{fig1}b). This requires atomic-scale magnetic control, which, however, is impracticable for conventional magnetic techniques. The recent advent of 2D van der Waals layered magnets sheds new light on this intractable problem \cite{Mak2019,Huang2020}. A unique advantage of van der Waals materials is the appearance of an additional degree of freedom regarding interlayer atomic registry, i.e. the vertical arrangement of atomic sites of adjacent layers \cite{Geim2013,Tong2017,Seyler2019,Tran2019,Jin2019,Alexeev2019,Song2019,Li2019,Chen2019,Enaldiev2020,Stern2021,Yasuda2021,Woods2021,Kennes2021,Andrei2021,Wilson2021}. When the AFM sublattice sits differently on a magnetic substrate layer, they would experience a distinct interlayer magnetic interaction. In the Landau-Lifshitz theory \cite{Landau1935}, this sublattice dependent magnetic interaction defines an effective staggered sublattice magnetic field that operating at atomic scale  (c.f. Figure \ref{fig1}c). This staggered effective field arising from the unique stacking degree of freedom in van der Waals materials has not been considered in conventional magnetic control schemes, such as exchange bias where AFM materials are mainly used as passive pinning substrates \cite{Berkowitz1999}. This motivates us to explore the possibility to control AFM order in van der Waals layered 2D magnets in pursuit of tunable strong magnetic anisotropy.

Here, via stacking an AFM monolayer on a ferromagnetic (FM) substrate, we show that the interlayer magnetic interaction introduces a staggered sublattice magnetic field, which can be divided into an effective anisotropic field that stabilizes the AFM order and an effective Zeeman field that breaks the degeneracy of magnon modes. Interestingly, the effective anisotropic field couples strongly with interlayer stacking order and magnetic order of the substrate layer, and even can change sign that flips the AFM order. These two types of couplings affect distinctly AFM magnetization dynamics, where an interlayer shift changes the ratio of sublattice precession amplitudes while a magnetic field changes its chirality. In the moir\'{e} superlattice formed by a relative twist or strain between the magnetic layers, the stacking dependent effective anisotropic field creates nonuniform magnetic textures, including AFM skyrmions. The low-energy magnons feel a trapping potential contributed from both the AFM texture and the landscape of the effective field, and are periodically trapped in the moir\'{e}. Our findings, supported by first-principles calculations on MnPS$_3$/CrCl$_3$ heterobilayer, are general in layered 2D magnets (AFM MnPS$_3$ homobilayer is studied in the discussion section).

\section{\color{blue}{Effective magnetic anisotropy and its tunability}}\label{sec:system}
We first give a general analysis on how to create a tunable magnetic anisotropy from the sublattice dependent interlayer magnetic interaction. The model Hamiltonian for a commensurate bilayer consisting of an AFM top layer and an FM bottom layer is
\begin{eqnarray}
H^{bi}=\sum_{\langle i,j \rangle,l}(J^{l}\bm{m}^{l}_{i}\bm{m}^{l}_{j}-K^l_0 m^{2}_{z,i})-\sum_{i,j}J^{\perp}_{i,j}(\bm{r})\bm{m}^{t}_{i}\bm{m}^{b}_{j}
\end{eqnarray}
where $\bm{m}^{l}_{i}=\bm{M}^{l}_{i}/M^{l}_0$ is the normalized magnetic moment at $i$-site of the $l$-th layer with $M^{l}_0$ being its magnitude and $J^{l}$ ($J^{\perp}_{i,j}$) is the intralayer (interlayer) exchange coupling. The interlayer part depends on the interlayer distance and inplane shift $\bm{r}$ between the magnetic atoms of the two layers \cite{Tong2017}. The summation of lattice sites includes the sublattice degree of freedom and runs over nearest neighbor sites $\langle i,j \rangle$ in the intralayer part. $K_0^l$ is anisotropy energy, which is assumed to be small for the AFM top layer so that its magnetic order can be tuned.

In the Landau-Lifshitz theory, the magnetic moments of the top AFM layer at A and B sublattice feel a local effective field $\bm{H}^{eff}_{A/B}(\bm{r})=-\partial H^{bi}/\partial\bm{m}_{i,A/B}$. Neglecting the exchange field that does not affect the magnetic anisotropy and assuming the FM order in the bottom layer is uniform, i.e. $\bm{m}^{b}_{j}=\bm{n}$, this field reads
\begin{eqnarray}
\bm{H}^{eff}_{A/B}(\bm{r})&=&\pm\tau_0 K^t_0 \bm{z}+[\pm\tau K_{eff}(\bm{r})+Z_{eff}(\bm{r})]\bm{n}, \label{Heff}
\end{eqnarray}
where $\tau_0=sgn[\bm{l}\cdot\bm{z}]$ and $\tau=sgn[\bm{l}\cdot\bm{n}]$ with the AFM N\'{e}el vector $\bm{l}=(\bm{m}_A-\bm{m}_B)/2$, and we have defined $K_{eff}(\bm{r})=\sum_{j}[J^{\perp}_{j}(\bm{r}_{A})-J^{\perp}_{j}(\bm{r}_{B})]/2$ and $Z_{eff}(\bm{r})=\sum_{j}[J^{\perp}_{j}(\bm{r}_{A})+J^{\perp}_{j}(\bm{r}_{B})]/2$. From Eq. (\ref{Heff}), one immediately recognizes that the discrepancy of the two sublattice effective field $K_{eff}(\bm{r})$ acts as an effective anisotropic energy that can stabilize the AFM order. While their average $Z_{eff}(\bm{r})$ acts as a uniform effective Zeeman energy that can break the degeneracy of the AFM resonance modes and even lead to spin flop phase when sufficiently large. Importantly, contrary to the original anisotropic term $K_0^t$ that is fixed by the material, this effective anisotropic energy $K_{eff}(\bm{r})$, depending on not only the orientation of the FM order $\bm{n}$ but also on the relative distance $\bm{r}$ between the magnetic moments of the top and bottom layers, is highly tunable. With the former being controlled by an applied magnetic field and the latter by an interlayer relative shift \cite{Jiang20182}, these couplings provide unique possibilities to manipulate AFM order.

\begin{figure}
\centering\includegraphics[width=0.5\textwidth]{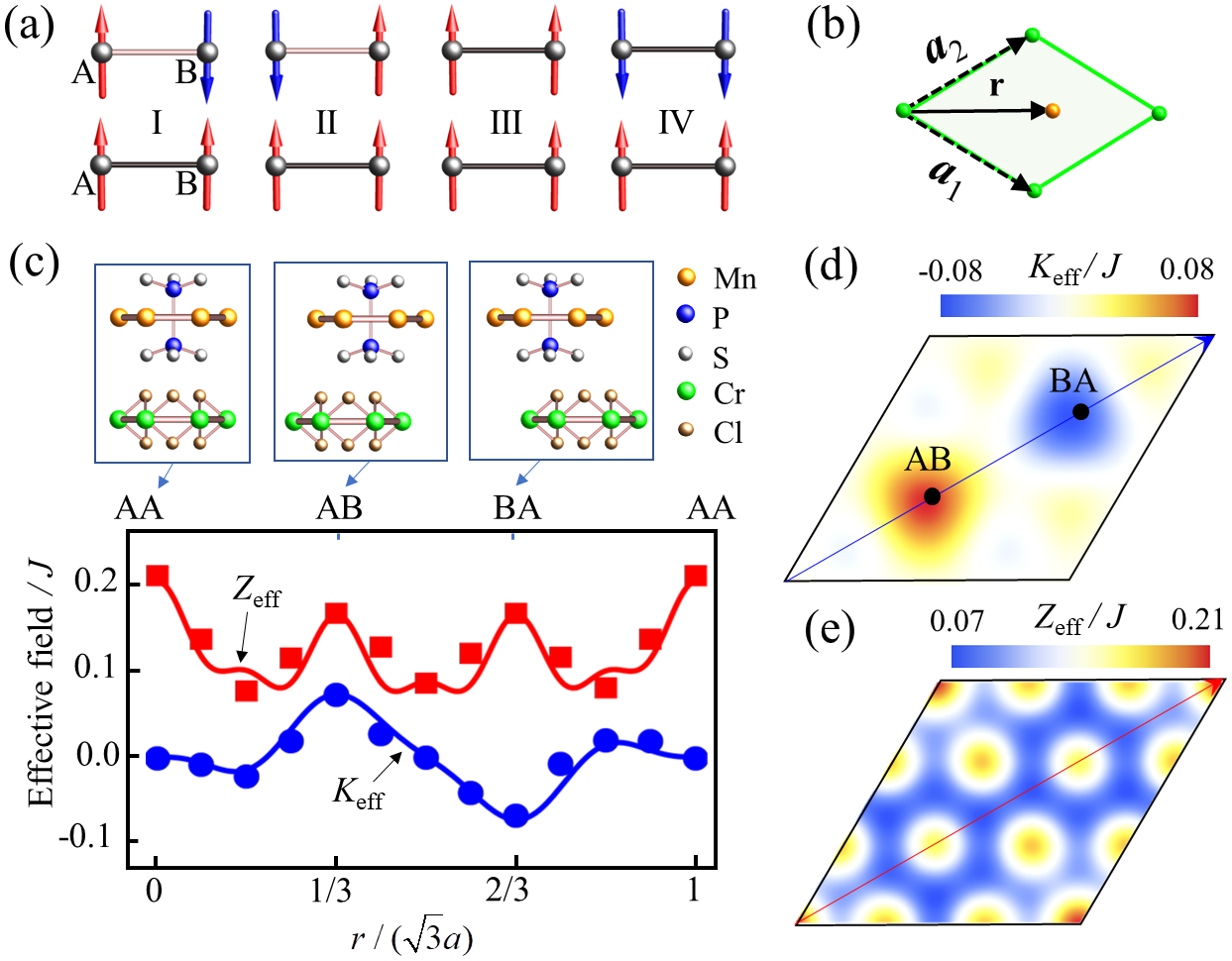}\caption{(a) Four magnetic configurations used to calculate the interlayer interaction induced effective fields. (b) A shifting vector $\bm{r}$ that characterizes interlayer atomic registry of a commensurate MnPS$_3$/CrCl$_3$ bilayer, which is defined in a unit cell with unit vectors $\bm{a}_{1/2}$. (c) First-principles calculated effective anisotropic field $K_{eff}$ (blue dots) and Zeeman field $Z_{eff}$ (red squares) in an AFM monolayer MnPS$_3$ induced by interlayer interaction from an FM monolayer CrCl$_3$ for various stacking registries along the long-diagonal in (b). Three high-symmetry configurations are shown atop. The curves are plotted using Eqs. (S1) and (S2) in Supporting Information 1, whose stacking dependence in a full unit cell are shown in (d) and (e). $\bm{n}=\bm{z}$ is used in the CrCl$_3$ layer. 
\label{fig2}}
\end{figure}

The underlying physics of the interlayer interaction induced effective anisotropy becomes more evident when one considers the AFM sublattice precession dynamics, via generalizing directly Kittel's original derivation in the case of a uniform magnetic field to a staggered sublattice magnetic field \cite{Keffer1952}. For an FM substrate with $\bm{n}=\bm{z}$, the AFM resonance frequency and the ratio of the precession amplitudes between two sublattice are (c.f. Supporting
Information 2)
\begin{eqnarray}
\omega_{\pm}(\bm{r})&=& \gamma Z_{eff}(\bm{r})\pm \gamma\{[K_0+\tau K_{eff}(\bm{r})] \nonumber \\
&\;&[2J+K_0+\tau K_{eff}(\bm{r})]\}^{1/2}, \nonumber \\
\chi_{\pm}(\bm{r})&=&-1-[K_0+\tau K_{eff}(\bm{r})]/J \mp \{[K_0+\tau K_{eff}(\bm{r})] \nonumber \\
&\;&[2J+K_0+\tau K_{eff}(\bm{r})]\}^{1/2}/J, \label{ratio}
\end{eqnarray}
where $\gamma$ is the gyromagnetic ratio and the superscript $t$ is omitted. These two ($\pm$) modes have opposite chiralities that the magnetization precesses circularly in opposite manners. It is evident that the effective anisotropic energy $K_{eff}(\bm{r})$ plays the same role as $K_0$ and can enhance the resonance frequency. Interestingly, with a typical small $K_0$, a stable AFM order requires $\tau K_{eff}(\bm{r})>0$, i.e. $K_{eff}(\bm{r})>0$ stables N\'{e}el order with $\tau=1$ and vice versa. The effective Zeeman field $Z_{eff}(\bm{r})$ lifts the degeneracy of the two modes, and the low-energy mode vanishes at a critical field value, beyond which spin flop phase develops.

The effective fields arising from the interlayer magnetic interaction can be quantitatively estimated from the energy discrepancy of certain magnetic configurations (c.f. Figure \ref{fig2}a). In particular, the effective anisotropic energy related to the discrepancy of the sublattice field can be calculated from the energy difference of configurations with top AFM layers that have opposite magnetic orders (I and II configurations in Figure \ref{fig2}a), which gives $E_{II}-E_I=4K_{eff}$. On the other hand, the effective Zeeman energy related to the average of the sublattice field can be measured by the one with top FM layers (III and IV configurations), which gives $E_{IV}-E_{III}=4Z_{eff}$.

In the following, we illustrate our results with a concrete example of a magnetic heterobilayer composed of an AFM MnPS$_3$ and an FM CrCl$_3$, both of which have been experimentally reported in its monolayer form and have almost identical honeycomb lattice constants. MnPS$_3$ has a very weak anisotropic energy $K_0\sim0.002J$ \cite{Okuda1986} and has been verified to be lack of long-range AFM order in its monolayer form \cite{Ni2021}. CrCl$_3$ crystal has weak in-plane anisotropy and its magnetic order can be tuned to out-of-plane by a small applied magnetic field, which is much weaker than the effective Zeeman field at the high-symmetry stackings \cite{McGuire2017}. For a commensurate honeycomb bilayer, the interlayer stacking order is characterized by a vector $\bm{r}$, which measures the inplane shift of a top Mn atom to a bottom Cr atom defined in a monolayer unit cell (c.f. Figure \ref{fig2}b).

Figure \ref{fig2}c shows the first-principles calculated effective anisotropic field (blue dots) and Zeeman field (red squares) along the long-diagonal of the unit cell, where there exist three high-symmetry configurations shown atop. At AB and BA stackings, where one of the AFM sublattice sits on top of the FM magnetic atom while the other is on top of the hollow center, the discrepancy of the interlayer magnetic interaction at the two sublattices is largest. Indeed, a strongest anisotropic field $K_{eff}$ is found, reaching as high as $40K_0$. Meanwhile, their opposite signs reflect a strong coupling between the anisotropic field and stacking order. At AA stacking, where the A and B sublattices of the top and bottom layer align, the effective field from the two sublattice matches, resulting in a zero effective anisotropic energy. Different from the anisotropic field, the effective Zeeman field is always positive that aligns with the substrate FM order, and approaches its maximum at AA stacking. Because the interlayer distance is much larger than the inplane atomic variation in a unit cell, one can well fit the datas using analytic expressions with low harmonic functions (c.f. curves in Figure \ref{fig2}c and see also Eqs. S(1) and S(2) in Supporting Information 1), whose variation in the whole unit cell are shown in Figures \ref{fig2}d and e \cite{Tong2017,Bistritzer2011}.

\begin{figure}
\centering\includegraphics[width=0.45\textwidth]{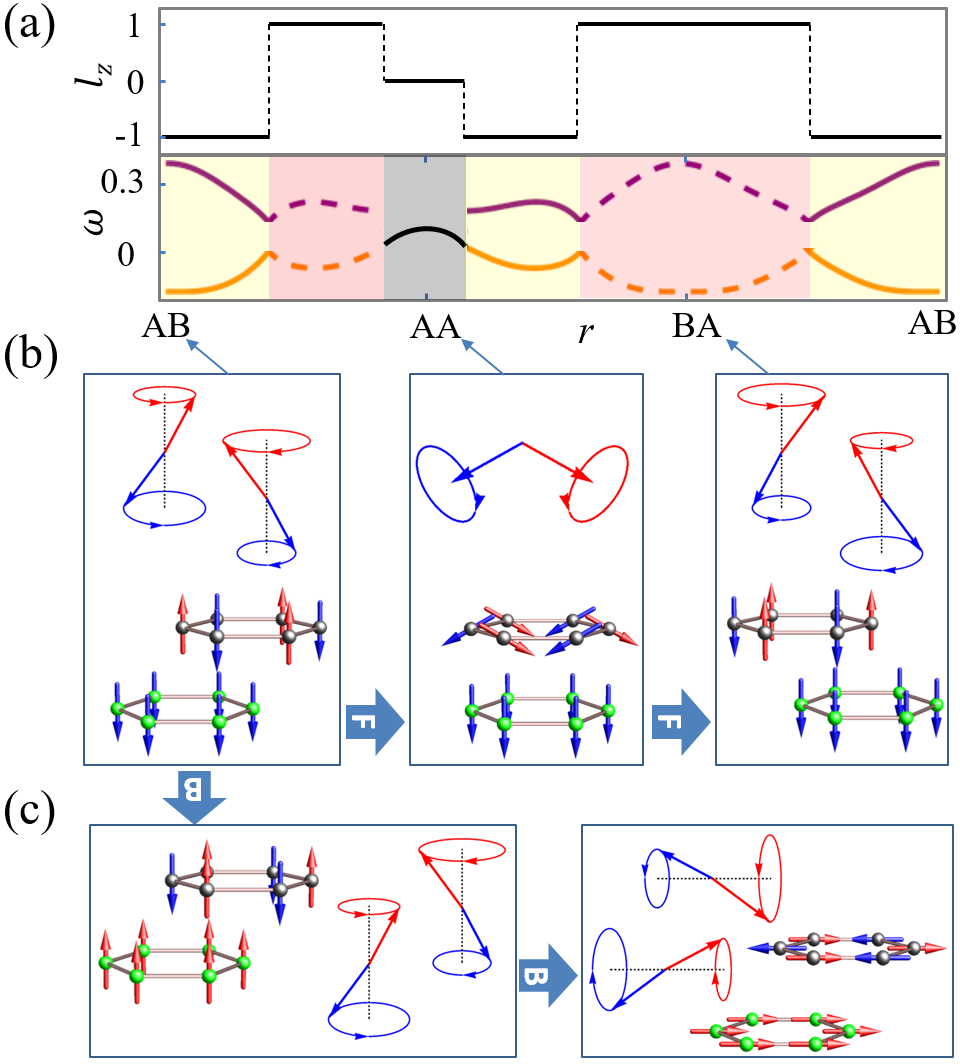}\caption{(a) Stacking dependence of $z$-component of the N\'{e}el order $l_z=(m^z_A-m^z_B)/2$ (top) and AFM resonance frequency $\omega$ (bottom). The N\'{e}el order is flipped around the AB and BA regions, and vanishes around the AA region that enters into spin flop phase. The effective fields are adopted from Figure \ref{fig2} with $\bm{n}=\bm{z}$. (b) Schematics of the precession of the AFM sublattice magnetization and its magnetic orientation for three high-symmetry configurations, which are related to each other via an interlayer shift between the layers exerted by a mechanical force (labeled by ``{\bf{F}}"). (c) A magnetic field (labeled by ``{\bf{B}}") tuning the substrate FM order changes the AFM order in the top layer and its dynamics.
\label{fig3}}
\end{figure}

The coupling between the effective anisotropy field and interlayer stacking order suggests a mechanical way to control the AFM order and its dynamics by shifting relatively the two layers. The top panel of Figure \ref{fig3}a shows the $z$-component of N\'{e}el order as a function of interlayer translation, when the effective fields lie out of plane. Near the AB and BA stackings, a uniform AFM magnetization with $|l_z|=1$ is developed but with an opposite sign, i.e. the N\'{e}el order flips, as schematically shown in Figure \ref{fig3}b. Near AA stacking, the strong Zeeman field together with weak anisotropy field destabilizes the AFM phase and favors a spin flop one with $l_z=0$. The lower panel of Figure \ref{fig3}a shows the stacking dependence of AFM resonance frequency, which is largest at AB and BA stackings, indicating their best magnetic stability. Because the effective Zeeman field is always positive, the lower-energy mode is always $\omega_{-}$ with a fixed chirality. At AB and BA stackings, although degenerate in energy, these two modes have opposite sublattice precession ratio because of the flipped N\'{e}el order (see Eq. (\ref{ratio})).

The coupling to the FM order of the substrate layer provides another magnetic way to control the AFM order and its dynamics (c.f. Figure \ref{fig3}c). Different from the mechanical control, which only flips the sign of the effective anisotropic field, reversing the FM order would flip the sign of the effective Zeeman field as well. In this case, the lower-energy mode changes from $\omega_{-}$ to $\omega_{+}$, which have opposite chiralities.

\begin{figure}[t]
\centering\includegraphics[width=0.45\textwidth]{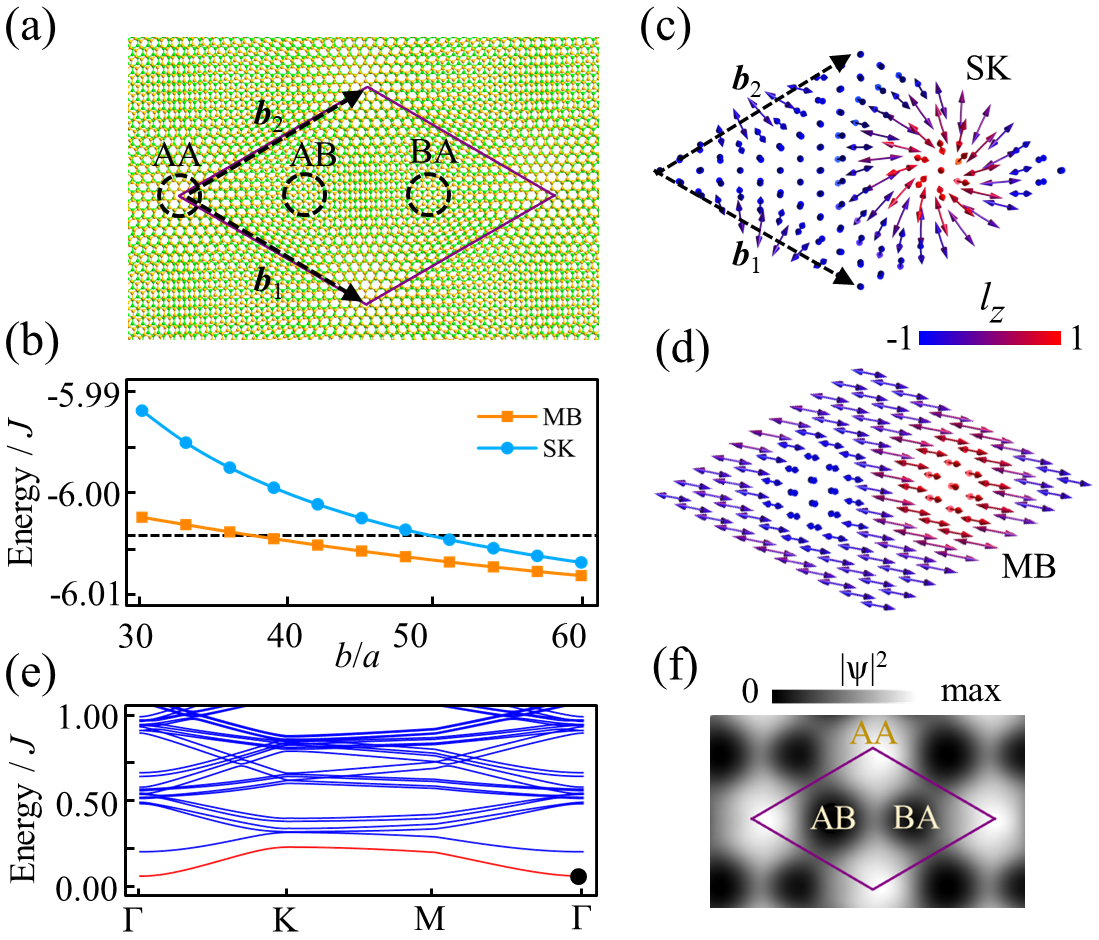}\caption{(a) Schematics of a moir\'{e} superlattice with unit vectors $\bm{b}_{1,2}$ formed by twisting a commensurate magnetic bilayer. (b) Energy per moir\'{e} cell as a function of moir\'{e} periodicity for AFM MB and SK textures, which are schematically shown in (c) and (d). The dashed line is the energy for uniform AFM state. The effective fields are adopted from Figure \ref{fig2} with $\bm{n}=\bm{z}$. (e) Magnon miniband of an MB texture with $b=42a$. (f) Wavefunction distribution of the lowest-energy magnon (black dot in (e)), which is localized around the AA region of a moir\'{e}.
\label{fig4}}
\end{figure}

\section{\color{blue}{AFM textures in magnetic moir\'{e} superlattices}}\label{sec:system}
The stacking dependent effective anisotropic field provides a unique possibility to engineer AFM textures \cite{Gomonay2018} in moir\'{e} superlattice formed by a relative twist or strain between the commensurate heterobilayer (c.f. Figure \ref{fig4}a). The periodicity $b$ of a moir\'{e} is inversely proportional to the misaligned angle and lattice mismatch, which can be further controlled by a twist or strain \cite{Ribeiro2018,Bai2020}. Compared with nonmagnetic one \cite{Geim2013,Tong2017,Seyler2019,Tran2019,Jin2019,Alexeev2019,Enaldiev2020,Stern2021,Yasuda2021,Woods2021}, a magnetic moir\'{e}, composed of lattices assigned with magnetic atoms, can be regarded as a superlattice of magnetic moments \cite{Tong2018,Sivadas20182,Hejazi2020,Akram2021,Song2021,Xu2021,Xie2021}. The smoothly changing atomic registry in a moir\'{e} then introduces a landscape of effective anisotropic and Zeeman fields for the top AFM layer. We assume that the superlattice is formed by two rigid lattices so that there exists a linear mapping between local region in the moir\'{e} and registry-dependent commensurate stacking defined in a monolayer unit cell. We notice that when the substrate FM layer has a nonuniform magnetic texture, the effective field would also show a spatial dependence even for a commensurate atomic bilayer. This FM texture is possible when a weak out-of-plane magnetic field is applied to tune the initially in-plane magnetic order in the CrCl$_3$ layer.

In terms of a spatially varying field, the AFM monolayer in the magnetic moir\'{e} can be modeled by an AFM monolayer imposed with laterally modulated anisotropic and Zeeman fields,
\begin{eqnarray}
H=\sum_{\langle i,j \rangle}J\bm{m}_{i}\bm{m}_{j}-[K_0+K_{eff}(\bm{R}_i)] m^{2}_{z,i}-Z_{eff}(\bm{R}_i) m_{z,i},\nonumber
\end{eqnarray}
where $\bm{R}_i$ is the atomic position in a moir\'{e} and we have assumed that the magnetization in the substrate layer is along $\bm{z}$-direction. The stable magnetic textures in a moir\'{e} can be obtained by relaxing from different initial magnetic configurations using Landau-Lifshitz-Gilbert equation \cite{Landau1935,Gilbert2004}. For a long-wavelength moir\'{e}, we find that the uniform AFM magnetization becomes unstable. Instead, magnetic textures, like AFM magnetic bubble (MB) and skyrmion (SK), become stable low-energy configurations (schematically shown in Figures \ref{fig4}c and d). With the increase of moir\'{e} periodicity, these AFM textures become more stable, as evidenced by the decrease in their energy (c.f. Figure \ref{fig4}b). Because SK has an inplane vortex structure, it costs more intralayer exchange energy and thus is energetically higher than MB. Interestingly, because these AFM textures are generated and stabilized by a landscape of the effective field, they can be driven by a slight interlayer shift and regulated by an external magnetic field, which provides new means to tune AFM textures (c.f. Supporting Information 3). These moir\'{e} generated AFM textures avoid the stringent requirement of Dzyaloshinskii-Moriya interaction \cite{Zhang2017}, which can be utilized to engineer topological phases of various quasiparticles \cite{Gobel2017,Diaz2019,Diaz2021}.

Usually the spatial modulation of the magnetic order in magnetic textures leads to low-energy magnons localized at domain walls \cite{Yu2021}. Instead, in moir\'{e} magnetic textures, there exists an additional spatially modulated effective field (c.f. Supporting Information 4). This field landscape tends to localize low-energy magnons around the AA region, where the magnetization is most unstable. Figure \ref{fig4}e shows the calculated magnon miniband, where a low-energy band is separated from others. Its wavefunction is localized around AA region of a moir\'{e} (c.f. Figure \ref{fig4}f). When thermal fluctuation is considered, magnetic orders in these regions are more fragile than the ones in other regions.

\section{\color{blue}{Discussion and conclusions}}
The effective anisotropic field that stabilizes AFM order arising from the interlayer magnetic interaction is a general result in van der Waals magnets. A recent experimental study on van der Waals AFM MnPS$_3$ has shown that long-range magnetic order persists from bulk to bilayer, while absent in monolayer \cite{Ni2021}. This is consistent with our result that interlayer magnetic interaction creates an effective anisotropy that stabilizes long-range magnetic order, while which is absent in the monolayer case without an interlayer interaction. For bilayer MnPS$_3$, our first-principles calculations show that the interlayer interaction induced effective anisotropic energy can reach as high as 12$K_0$  (c.f. Supporting Information 5).

In summary, we have revealed a tunable strong magnetic anisotropy in van der Waals magnets that can stabilize and manipulate 2D AFM order. This magnetic anisotropy arises from a sublattice dependent interlayer magnetic interaction that strongly couples with the interlayer stacking order and magnetic order of the substrate layer. These couplings provide unprecedented mechanical and magnetic controls over AFM order as well as its dynamics. In the moir\'{e} of a magnetic heterostructure, the stacking dependent anisotropic field can be utilized to engineer nonuniform AFM textures, in which low-energy magnons are periodically trapped in the moir\'{e}. Our results point to a new route to explore and manipulate 2D AFM magnetism, with potential applications in constructing atomic-scale AFM spintronic devices and studying magnon condensation physics.

\section{\color{blue}{Supporting Information}}
The Supporting Information includes details of the first-principles calculations, AFM magnetization dynamics, mechanical and magnetic control over the moir\'{e} AFM textures, localized magnons in moir\'{e} AFM textures, and effective anisotropic field in bilayer MnPS$_3$.

\section{\color{blue}{Acknowledgments}}
We thank Wang Yao for stimulating discussions at initial stage of this work. This work is supported by the National Natural Science Foundation of China (11904095), the National Key Research and Development Program of Ministry of Science and Technology (2021YFA1200503), and the Fundamental Research Funds for the Central Universities from China.

\end{document}